\documentclass[journal,twoside,web]{ieeecolor}
\usepackage{tmi}
\usepackage{cite}
\usepackage{amsmath,amssymb,amsfonts}
\usepackage{algorithmic}
\usepackage{graphicx}
\usepackage{textcomp}
\usepackage{multirow}
\usepackage{booktabs}
\usepackage[table]{xcolor}
\usepackage[table,dvipsnames]{xcolor}
\usepackage{float}

\usepackage[utf8]{inputenc}
\usepackage[linesnumbered,ruled,vlined]{algorithm2e}
\usepackage{url}
\usepackage{pifont}
\newcommand{\xmark}{\ding{55}}
\usepackage{changes}

\def\BibTeX{{\rm B\kern-.05em{\sc i\kern-.025em b}\kern-.08em
    T\kern-.1667em\lower.7ex\hbox{E}\kern-.125emX}}
\begin{document}
\title{Accelerating Volumetric Medical Image Annotation via Short-Long Memory SAM 2}
\author{Yuwen Chen, Zafer Yildiz, Qihang Li, Yaqian Chen, Haoyu Dong, Hanxue Gu, Nicholas Konz, Maciej A. Mazurowski
\thanks{Manuscript submitted on May 3 and accepted on October 27. This work was supported by the National Institute Of Biomedical Imaging And Bioengineering of the National Institutes of Health under Award Number R01EB031575.}
\thanks{Yuwen Chen, Yaqian Chen, Haoyu Dong, Hanxue Gu, Nicholas Konz are with the Department of Electrical and Computer Engineering, Duke University, Durham, NC 27708 USA (e-mail: yuwen.chen@duke.edu; yaqian.chen@duke.edu; haoyu.dong151@duke.edu; hanxue.gu@duke.edu; nicholas.konz@duke.edu)}
\thanks{Zafer Yildiz and Qihang Li are with the Department of Biostatistics and Bioinformatics, Duke University, Durham, NC 27708 USA (e-mail: zafer.yildiz@duke.edu; qihang.li@duke.edu)}
\thanks{Maciej A. Mazurowski is with Department of Biostatistics and Bioinformatics, Radiology, Electrical and Computer Engineering, and Computer Science, Duke University, Durham, NC 27708 USA (e-mail: maciej.mazurowski@duke.edu)}
}

\maketitle

\begin{abstract}
Manual annotation of volumetric medical images, such as magnetic resonance imaging (MRI) and computed tomography (CT), is a labor-intensive and time-consuming process. Recent advancements in foundation models for video object segmentation, such as Segment Anything Model 2 (SAM 2), offer a potential opportunity to significantly speed up the annotation process by manually annotating one or a few slices and then propagating target masks across the entire volume. However, the performance of SAM 2 in this context varies. Our experiments show that relying on a single memory bank and attention module is prone to error propagation, particularly at boundary regions where the target is present in the previous slice but absent in the current one. To address this problem, we propose Short-Long Memory SAM 2 (SLM-SAM 2), a novel architecture that integrates distinct short-term and long-term memory banks with separate attention modules to improve segmentation accuracy. We evaluate SLM-SAM 2 on four public datasets covering organs, bones, and muscles across MRI, CT, and ultrasound videos. We show that the proposed method markedly outperforms the default SAM 2, achieving an average Dice Similarity Coefficient improvement of 0.14 and 0.10 in the scenarios when 5 volumes and 1 volume are available for the initial adaptation, respectively. SLM-SAM 2 also exhibits stronger resistance to over-propagation, reducing the time required to correct propagated masks by 60.575$\%$ per volume compared to SAM 2, making a notable step toward more accurate automated annotation of medical images for segmentation model development.
\end{abstract}


\section{Introduction}
\label{sec:introduction}
\IEEEPARstart{S}{emantic} segmentation plays a crucial role in medical image analysis, serving as a foundational step in numerous clinical and research applications, including tumor identification and measurement, treatment planning, and body composition analysis \cite{weston2019automated,higgins2021machine, lee2021deep,gu2025segmentanybone}. With advances in computer vision, deep learning-based methods (e.g., UNet \cite{unet}) have increasingly become the dominant approach for medical image segmentation due to their superior robustness and efficiency. However, developing reliable deep learning models often demands high-quality manual annotations, which are time-consuming and expensive to acquire \cite{aljabri2022towards}, particularly for 3-dimensional (3D) volumetric medical data (e.g., MRI and CT) and medical videos (e.g., ultrasound videos) that consist of numerous 2D slices, requiring detailed labeling.

3D medical imaging data typically consists of a stack of 2D slices acquired along an additional spatial axis. These 2D consecutive slices are often highly correlated, capturing gradual anatomical changes across the axis. Consequently, the 3D volume data can be interpreted as a temporal progression, analogous to frames in a video \cite{cheng2025interactive}. Leveraging such similarity between adjacent slices offers an effective method to accelerate the annotation process by propagating annotations from a single slice to the rest of the volume \cite{yeung2021sli2vol,bitarafan2022vol2flow,wu2022self,an2025sli2vol+}.

Several approaches have employed self-supervised learning (e.g., training registration networks) to learn semantic correspondences between adjacent slices. These correspondences are subsequently utilized to propagate annotations from a labeled slice to neighboring slices for segmentation \cite{yeung2021sli2vol,bitarafan2022vol2flow,wu2022self,an2025sli2vol+}. However, these methods are prone to \textit{error drift}, where errors accumulate during propagation. Furthermore, they exhibit limitations in handling discontinuities within 3D volumes, such as the emergence of previously unseen anatomical structures or the disappearance of existing ones \cite{an2025sli2vol+}.

The recently developed video object segmentation (VOS) foundation model, Segment Anything Model 2 (SAM 2) \cite{sam2}, has been applied to medical imaging segmentation \cite{wu2025medical,cheng2025interactive}, offering an alternative. Vision foundation models have demonstrated exceptional generalization capabilities without requiring access to large-scale labeled datasets \cite{kirillov2023segment,mazurowski2023segment,gu2025segmentanybone}. Through extensive pretraining, these models acquire robust and adaptable feature representations, which facilitate their applications to downstream tasks with limited annotated data \cite{gu2025segmentanybone}. In VOS tasks, the objective is to track objects throughout a video sequence by propagating an initial object mask provided in the first frame \cite{yao2020video,sam2}. By interpreting the depth dimension as the temporal axis, SAM 2 offers a supervised method (i.e., training with images and corresponding masks) to propagate annotation masks across slices. For instance, SAM 2 has been utilized to build a 3D interaction framework \cite{cheng2025interactive}.

Based on SAM \cite{kirillov2023segment}, SAM 2 enables video object segmentation by introducing a memory encoder, memory bank, and memory attention \cite{sam2}. The memory encoder constructs memory representation through element-wise summation of encoded images and their corresponding output masks. The memory bank serves as a repository that retains information and past predictions from $\boldsymbol{N}$ recent slices and $\boldsymbol{M}$ prompted slices. Additionally, the memory attention module enhances the feature representation of the current slice by conditioning it on the features and predictions of past slices from the memory bank. The output features from the memory attention are then input to the mask decoder to generate a mask prediction.

Nevertheless, our experiments (Sec. \ref{sec:results}) reveal that SAM 2, under its default settings ($N+M=7$), exhibits suboptimal performance at discontinuous slice boundaries, particularly during object disappearance or emergence. The limitations can be attributed to two primary factors: (1) persistent retention of prompted slices in the memory bank throughout the entire propagation, which can introduce outdated or irrelevant context, and (2) reliance on multiple recent slices ($N \ge 5$), which may result in over-segmentation due to the accumulation of momentum from prior frames. These issues highlight the need for a more adaptive memory mechanism that can effectively keep temporal consistency while maintaining sufficient flexibility to address abrupt changes and discontinuities.

Intuitively, a longer memory (i.e., larger $N$) allows the model to track gradual changes in the object, thereby enhancing the stability and continuity of predictions over slices. In contrast, a shorter memory (i.e., smaller $N$) emphasizes the most recent slices, improving the model's adaptability to discontinuous changes in object appearance. To leverage the complementary advantages of both long-term and short-term memory, we propose \textbf{S}hort-\textbf{L}ong \textbf{M}emory SAM 2 (\textbf{SLM-SAM 2}) \footnote{GitHub at: \url{https://github.com/mazurowski-lab/SLM-SAM2}}, a novel framework that incorporates two memory banks, one long-term and one short-term, to achieve more robust and accurate segmentation. We validate the effectiveness of SLM-SAM 2 through experiments on four public datasets spanning MRI, CT, and ultrasound video modalities. The results demonstrate that SLM-SAM 2 significantly outperforms existing 2D and 3D segmentation methods. Our contributions are summarized below:

\begin{itemize}
    \item We propose a novel architecture, SLM-SAM 2, which leverages the complementary strengths of long-term and short-term memory to facilitate and enhance 3D medical image annotations.
    \item Our method significantly outperforms existing approaches in segmenting organs, bones, and muscles across MRI, CT, and ultrasound videos, achieving an average dice score improvement of at least $7\%$ across different datasets.
    \item We demonstrate both qualitatively and quantitatively that SLM-SAM 2 can effectively alleviate the over-propagation issue and other common propagation errors.
    \item Our method significantly reduces the manual efforts in correcting propagated masks by $60.575\%$ in seconds per volume and $53.574\%$ in corrected-slice ratio compared to SAM 2.
    \item We integrate SLM-SAM 2 into 3D Slicer and release the extension to facilitate 3D medical data annotation.
    
\end{itemize}

\section{Related Work}
\label{sec:relatedwork}

\subsection{Semantic Segmentation in Medical Imaging}
\label{sec:semantic_segmentation}
Semantic segmentation aims to classify each pixel in an image, making it an essential task in both natural and medical imaging \cite{asgari2021deep,xu2024advances}. Convolutional Neural Networks (CNNs) have been a dominant approach for vision tasks, with UNet \cite{unet} emerging as a widely adopted model in medical image segmentation. The skip connections between the encoder and decoder allow UNet to effectively capture both global and local context. Building on UNet, nnUNet \cite{nnunet} introduces a fully automated training pipeline configuration, which can be efficiently adapted to any new medical segmentation tasks. Nevertheless, one of the key limitations of CNNs is that they often struggle to capture long-range dependencies \cite{kim2025systematic}.

To overcome this issue, Vision Transformers (ViTs) employ a transformer architecture that divides images into a sequence of patches \cite{dosovitskiyimage}. The self-attention mechanism in ViTs allows them to capture global context, enhancing segmentation accuracy for anatomies with large receptive fields (e.g., the liver). Recently, vision foundation models, such as SAM \cite{ma2024segment}, have further advanced the medical image segmentation. SAM leverages a transformer-based encoder and extensive pretraining on large-scale datasets, enabling exceptional generalization capabilities across multiple downstream tasks with minimal labeled data \cite{wang2024fsam,wu2025medical}.

Recurrent networks, such as convolutional Long Short-Term Memory (ConvLSTM) and its variants, have also been explored in medical image segmentation \cite{chen2016combining,novikov2018deep,azad2019bi,xu2019lstm}. For instance, \cite{azad2019bi} replaced skip connection of UNet with Bidirectional ConvLSTM \cite{song2018pyramid} to enhance intra-slice feature fusion. More related, \cite{novikov2018deep} combined Bidirectional ConvLSTM with UNet to process full volumes segmentation sequentially, capturing dependencies across slices. Similarly, \cite{xu2019lstm} introduced an LSTM multi-modal UNet to enforce inter-slice continuity for brain tumor segmentation. Nevertheless, these approaches rely on implicit hidden-state propagation to model inter-slice context and often lack object-specific memory mechanisms, limiting their capacity for fine-grained segmentation and robustness over long sequences.

\begin{figure*}[!h]
    \centering
    \includegraphics[width=1\linewidth]{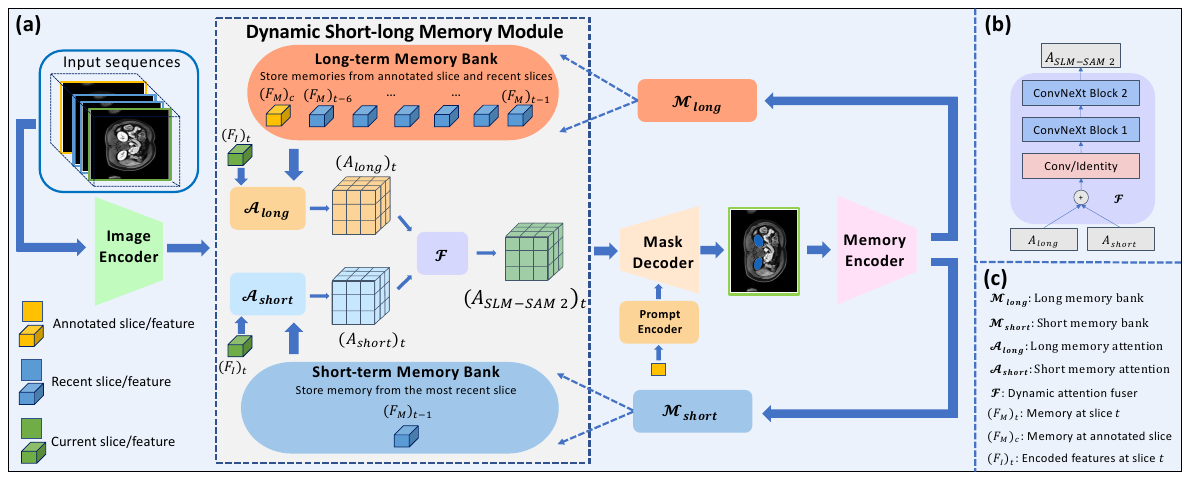}
    \caption{\textbf{Overview of SLM-SAM 2.} (a) is the general pipeline of SLM-SAM 2. The dynamic short-long memory module consists of short-term and long-term memory banks ($\mathcal{M}_{short}$ and $\mathcal{M}_{long}$), two distinct attention modules for each of them ($\mathcal{A}_{short}$ and $\mathcal{A}_{long}$), and the dynamic attention fuser ($\mathcal{F}$). Specifically, $\mathcal{M}_{long}$ includes memories from one annotated slice and up to six additional slices, while $\mathcal{M}_{short}$ contains memory only from the most recent slice. $A_{short}$ and $A_{long}$ denote the outputs of $\mathcal{A}_{short}$ and $\mathcal{A}_{long}$, respectively, and $A_{SLM-SAM 2}$ represents the fused attention features produced by $\mathcal{F}$. (b) illustrates the architecture of the dynamic attention fuser $\mathcal{F}$. (c) provides an explanation of the key notations used in SLM-SAM 2.}
    \label{fig:pipeline}
\end{figure*}

\subsection{Video Object Segmentation (VOS)}
\label{sec:video_object_segmentation}
Video object segmentation focuses on segmenting particular object instance given an entire video sequence and an object mask on the first frame, either manually annotated or automatically generated \cite{yao2020video}. VOS has numerous applications in various domains, including autonomous vehicles \cite{siam2021video,kozlowski2024image}, robotics \cite{miao2024region,manakitsa2024review}, and medical imaging analysis \cite{cheng2025interactive,min2025innovating}. VOS methods can be broadly categorized into unsupervised VOS and semi-supervised VOS \cite{gao2023deep}. Unsupervised VOS aims to identify and segment the most salient objects in a video without any given annotations. In contrast, semi-supervised VOS requires initial annotations, which are typically provided in the first frame \cite{zhou2022survey}. In addition, semi-supervised VOS methods also can be adopted for interactive VOS segmentation by combining with SAM \cite{cheng2023tracking,rajivc2025segment} or other 2D interactive segmentation methods \cite{isegformer}.

Memory-based architectures have been widely adopted in advancing semi-supervised VOS. For example, STM \cite{oh2019video} introduces a space-time memory network that stores past image-mask pairs in a single memory bank, retrieving them via a key-value attention mechanism. XMem \cite{cheng2022xmem} extends this paradigm with an Atkinson-Shiffrin-inspired design that maintains multiple memory stores for both long-term and short-term information, and consolidates frequently used prototypes based on cumulative affinity scores. Recently, additional approaches, such as Cutie \cite{cutie} and SAM 2 \cite{sam2}, have been introduced for semi-supervised VOS tasks. Cutie employs an iterative interaction between top-down object-level memory reading and bottom-up pixel-level features to effectively capture the semantic structure of foreground objects \cite{cutie}. SAM 2 incorporates a pretrained ViT backbone along with a memory mechanism that includes a memory bank, memory encoder, and memory attention module \cite{sam2}. The enriched feature representations from the pretrained ViT encoder allow SAM 2 to achieve robust generalization performance across different tasks. Different from previous interactive VOS approaches, SAM 2 offers a unified model that directly accepts user prompts for real-time mask refinement in videos. Inspired by these prior works, our method builds on SAM 2 by integrating dual short-term and long-term memory banks and introducing a learnable fuser that dynamically combines the two memories. This design provides a novel mechanism to leverage SAM 2's rich encoder and flexible prompting while enabling adaptive use of memory information for efficient volumetric medical image annotation.

\subsection{Medical Image Annotation for Segmentation}
\label{sec:medical_image_annotation}
Medical image annotation for segmentation tasks involves pixel-level labeling of images across various modalities to train deep learning-based segmentation models\cite{aljabri2022towards}. The manual annotation process is both expensive and time-consuming due to the complexity of different modalities and anatomical structures \cite{wang2021annotation,bonaldi2023deep}. Moreover, a significant portion of medical images is acquired in three-dimensional formats, including MRI, CT, and video-based imaging (e.g., colonoscopy and ultrasound), further increasing the annotation workload.

To address these challenges, various approaches have been developed, which can be broadly categorized into unsupervised, supervised, and interactive methods. Unsupervised methods focus on learning spatial relationship between adjacent slices in 3D data, enabling label propagation from a single annotated slice to the entire volume during inference \cite{yeung2021sli2vol,an2025sli2vol+}. However, these methods are prone to error drift \cite{an2025sli2vol+}. Supervised methods often utilize few-shot learning \cite{feng2021interactive,pachetti2024systematic}, where a model is trained on a small, fully annotated subset of data before being applied to the remaining data. This category involves 2D \cite{xie2024sam,ma2024segment}, 3D \cite{unetr,swinunetr}, and VOS models \cite{cutie,sam2}, with VOS technique sometimes requiring minimal annotations during the inference stage. Interactive annotation methods integrate automated predictions with real-time human corrections, enabling quick refinement of initial annotations, and thus improving annotation efficiency \cite{feng2021interactive}.
\section{Methods}
\label{sec:methods}
This work aims to develop robust and reliable slice propagation models that can accelerate the annotation of volumetric medical image data. In this section, we define the problem and introduce the proposed SLM-SAM 2 in detail.

\subsection{Problem Definition}
\label{sec:problem_definition}
Given a fully annotated volume set $\{X_i|i=0, ..., n\}$ for training, the objective is to segment any incoming volumes from $\{Y_i|i=0, ...,m\}$ by annotating a single slice from each volume. In this study, we chose $n=5$, reflecting a practical setting in which users initially provide full annotations for a limited number of volumes and subsequently annotate only one slice for each newly acquired volume.

\subsection{Model Architecture of SLM-SAM 2}
\label{sec:model_architecture}
Fig. \ref{fig:pipeline} (a) depicts the overall SLM-SAM 2 pipeline. SLM-SAM 2 retains the foundational architecture of SAM 2, including the image encoder $E_{I}$, prompt encoder $E_{P}$, and mask decoder $Dec$, but replaces its single memory bank and attention module with a novel dynamic short-long memory module, as shown in Fig. \ref{fig:pipeline}.

Given an input sequence of frames, the image encoder of SLM-SAM 2 produces a feature embedding of each frame. The initial prompt (e.g., manual annotations) is encoded by the prompt encoder. For mask prediction, each feature embedding is dynamically conditioned on short-term and long-term memories via separate learnable attention modules and then fused by a dynamic, trainable fusion module. The resulting embeddings are passed to the mask decoder to generate segmentation mask. The memory of each frame is generated from the memory encoder using its embedding features from the image encoder together with the corresponding predicted mask.

SLM-SAM 2 adopts the ``tiny'' Hiera image encoder \cite{ryali2023hiera} as the backbone, which is pretrained using MAE \cite{he2022masked}. The prompt encoder of SLM-SAM 2 follows the same architecture of SAM 2 to allow mask inputs, and SLM-SAM 2 employs the same mask decoder architecture as SAM 2. The key dynamic short-long memory module consists of a short-long memory bank mechanism and a dynamic attention fuser, as described below.

\subsubsection{Short-long Memory Bank Mechanism}
\begin{algorithm}[!h]
\caption{SLM-SAM 2 Training Pseudo-Code}
\label{alg:sam2_train}
\KwIn{\\
  \quad $I_{\text{seq}} = \{ I_1, I_2, \ldots, I_T \}$: Sequence of $T$ frames\\
  \quad $P$: Mask prompt\\
  \quad $D = \{ (I_{\text{seq}}, P, S^{gt}_{\text{seq}}) \}$: Training data sequences\\
  \quad $S^{gt}_{\text{seq}} = \{ S^{gt}_1, S^{gt}_2, \ldots, S^{gt}_T \}$: Ground-truth masks\\
  \quad $N$: Number of training epochs.\\
  \quad $\mathcal{M}_{long}$: Long-term memory bank\\
  \quad $\mathcal{M}_{short}$: Short-term memory bank\\
  \quad $\mathcal{A}_{long}$: Long-term memory attention module\\
  \quad $\mathcal{A}_{short}$: Short-term memory attention module\\
  \quad $\mathcal{F}$: Dynamic attention fuser\\
  \quad $E_{P}, \boldsymbol{F}_P$: Prompt encoder and its output\\
  \quad $E_{I}, \boldsymbol{F}_I$: Image encoder and its output\\
  \quad $E_{M}, \boldsymbol{F}_M$: Memory encoder and its output\\
  \quad $Dec$: Mask decoder\\
  \quad $S_{t}^{pred}$: Predicted mask at slice $t$\\
  
}
\KwOut{Trained model parameters.}
\BlankLine
\For{$\text{epoch} \gets 1$ \KwTo $N$}{
  \For{each $(I_{\text{seq}}, P, S^{gt}_{\text{seq}})$ in $D$}{
    $\mathcal{M}_{long} \gets \varnothing$ \tcp*[r]{long-term memory}
    $\mathcal{M}_{short} \gets \varnothing$ \tcp*[r]{short-term memory}
    $\boldsymbol{F}_P \gets E_{P}(P)$\;
    $(\boldsymbol{F}_M)_0 \gets \boldsymbol{0}$\;
    $L_{\text{total}} \gets 0$ \tcp*[r]{total loss}\
    \For{$t \gets 1$ \KwTo $T$}{
      \BlankLine
      \tcp{Forward pass}
      $(\boldsymbol{F}_I)_t \gets E_{I}(I_t)$\;
      $(A_{long})_t \gets \mathcal{A}_{long}((\boldsymbol{F}_I)_t, \mathcal{M}_{long})$\;
      $(A_{short})_t \gets \mathcal{A}_{short}((\boldsymbol{F}_I)_t, \mathcal{M}_{short})$\;
      $(A_{SLM-SAM 2})_t \gets \mathcal{F}((A_{long})_t, (A_{short})_t)$\;
      
      \BlankLine
      \tcp{Prediction at slice $t$}
      $S_t^{pred} \gets Dec((A_{SLM-SAM 2})_t, \boldsymbol{F}_P)$\;

      \BlankLine
      \tcp{Compute loss}
      Compute $L_{\text{step}}$\;
      $L_{\text{total}} \gets L_{\text{total}} + L_{\text{step}}$\;

      \BlankLine
      \tcp{Compute memories}
      $(\boldsymbol{F}_M)_t \gets E_{M}(I_t, S_t^{pred})$\;
      
      \BlankLine
      \tcp{Update memory}
      $\mathcal{M}_{long} \gets \textsc{Update}(\mathcal{M}_{long}, (\boldsymbol{F}_M)_t)$\;
      $\mathcal{M}_{short} \gets \textsc{Update}(\mathcal{M}_{short}, (\boldsymbol{F}_M)_t)$\;
    }

    \tcp{Backpropagate per sequence}
    \textsc{Reset\_Gradients()}\;
    \textsc{Backpropagate}$(L_{\text{total}})$\;
    \textsc{Optimizer\_Step()}\;
  }
}

\Return{Trained model parameters.}
\end{algorithm}

In SAM 2, at slice $I_t$, the memory attention $A_{SAM 2}$ can be expressed as \footnote{We denote slice, embedding, and memory at frame $t$ using $(.)_t$}:
\begin{equation}
    (A_{SAM 2})_t = \mathcal{A}((\boldsymbol{F}_{I})_t, \mathcal{M})
\end{equation}
\begin{equation}
    (\boldsymbol{F}_{I})_t = E_{I}(I_t)
\end{equation}
\begin{equation}
    \mathcal{M} = \{(\boldsymbol{F}_M)_j|(\boldsymbol{F}_M)_j \in \mathcal{M}\}
\end{equation}
where $(\boldsymbol{F}_I)_t$ is the output embedding feature from the image encoder at slice $I_t$, $(\boldsymbol{F}_M)_j$ represents memory at slice $I_j$, and $\mathcal{M}$ is the memory bank which stores these memories. The memory encoder uses a convolutional module to downsample the predicted mask and sum it element-wise with the corresponding image embedding, then applies a lightweight convolutional layer to produce the memory.

Similar to SAM 2, SLM-SAM 2 inputs the new incoming slices to the image encoder to generate the corresponding feature representations. To enable both short-term and long-term memory storage, SLM-SAM 2 introduces an additional memory bank and memory attention. The long-term memory bank $\mathcal{M}_{long}$ follows the original SAM 2 design by storing $N_{long}$ recent and $M_{long}$ prompted slices, ensuring continuity in segmentation. The short-term memory bank $\mathcal{M}_{short}$ solely retains the most recent $N_{short}$ slice, where $N_{short}$ is a typically small positive integer, allowing model predictions to quickly adapt to abrupt object changes. During propagation, the current slice's features attend independently to the short-term memory and the long-term memory, producing attention outputs $A_{short}$ and $A_{long}$.
\begin{equation}
    (A_{short})_t = \mathcal{A}_{short}((\boldsymbol{F}_{I})_t, \mathcal{M}_{short})
\end{equation}

\begin{equation}
    (A_{long})_t = \mathcal{A}_{long}((\boldsymbol{F}_{I})_t, \mathcal{M}_{long})
\end{equation}

The two memory attention modules are of the same architecture, resulting in output attentions of the same size. Each module consists of a stack of $L$ transformer blocks. Specifically, SLM-SAM 2 uses $N_{long}=6$, $M_{long}=1$ and $N_{short}=1$. Consequently, the contents of the two memory banks at slice $i$ can be described as:

\begin{equation}
    \mathcal{M}_{short} = \{(\boldsymbol{F}_{M})_{t-1}\}
\end{equation}

\begin{equation}
    \mathcal{M}_{long} = \{(\boldsymbol{F}_{M})_{t-1}, (\boldsymbol{F}_{M})_{t-2}, ..., (\boldsymbol{F}_{M})_{t-6}, (\boldsymbol{F}_{M})_{c}\}
\end{equation}

where $(\boldsymbol{F}_{M})_{c}$ is the memory on the conditional slice (i.e., the initial annotated slice).

Each memory bank operates as a FIFO queue of recent-slice memories. At each update (in Algorithm \ref{alg:sam2_train}), the short-term memory bank, $\mathcal{M}_{short}$, discards its current memory and enqueues the newly generated one. The long-term memory bank, $\mathcal{M}_{long}$, consistently retains the memory on conditional slices by appending the new memory but only dequeues its oldest entry on non-conditional slices.
  
\subsubsection{Dynamic Attention Fuser}
The two output attentions are of the same shape $(B, C, H, W)$ with $C=256, H=W=64$ and are subsequently fused using a lightweight, learnable fusion module $\mathcal{F}$. The fusion module first performs element-wise summation of $A_{short}$ and $A_{long}$, followed by an optional projection layer and two ConvNeXt blocks \cite{liu2022convnet}, as shown in Fig. \ref{fig:pipeline} (b). Each ConvNeXt begins with a depthwise $7 \times 7$ convolution followed by LayerNorm. The normalized output is then permuted to $(B, H, W, C)$ and passed through a two-layer pointwise MLP that expands channels from $C$ to $4C$, applies GELU, then projects back to $C$. Finally, the output is permuted back to $(B, C, H, W)$ and combined with the original input via a residual connection. Simply, $A_{SLM-SAM 2}$ can be computed as:

\begin{equation}
    (A_{SLM-SAM 2})_t = \mathcal{F}((A_{short})_t, (A_{long})_t)
\end{equation}

Thus, the fused attention $A_{SLM-SAM 2}$ is of the same shape as $A_{SAM 2}$, allowing it to be directly fed into the mask decoder to generate output predictions. All newly introduced modules adopt the same architecture as the original SAM 2 components to leverage the pretrained weights.

\subsection{Annotation Propagation via SLM-SAM 2}
\label{sec:annotation_propagation}
The finetuning process of SLM-SAM 2 is described in Algorithm \ref{alg:sam2_train}. The loss function incorporates focal loss, Dice loss, MAE loss for IoU prediction, and cross-entropy loss, with weights the same as those used in SAM 2 \cite{sam2}. At inference, the middle slice of the object of interest is selected as the initial annotation point, and the mask is propagated in a bi-directional manner, first toward one direction and subsequently in the opposite direction. To enhance the model's ability to track the object in both directions, we randomly flip half of the sequence order during training.

\section{Experiments}
\label{sec:experiment}

We have performed a thorough evaluation of the proposed SLM-SAM 2. In this section, we describe datasets, implementation details, and baselines, followed by results including ablation studies to verify the key designs of SLM-SAM 2, propagation error analysis, and quantitative efficiency experiments.

\subsection{Datasets}
\label{sec:datasets}
We evaluated SLM-SAM 2 on four publicly available datasets spanning different imaging modalities, covering anatomical structures including organs, bones, and muscles.
\subsubsection{Organ Segmentation} We used AMOS dataset \cite{amos} to segment the kidneys and pancreas in MRI scans (MRI-Kidney and MRI-Pancreas) and the pancreas and spleen in CT scans (CT-Pancreas and CT-Spleen), with 25 volumes randomly selected for each modality. Heart segmentation was performed on the CT dataset from the TotalSegmentator project \cite{totalsegmentator} (CT-Heart), using 35 randomly selected volumes for each task. 
\subsubsection{Bone Segmentation} We assessed CT femur segmentation task on randomly selected 35 volumes from the TotalSegmentator project \cite{totalsegmentator} (CT-Femur).
\subsubsection{Muscle Segmentation} The HuashanMyo MRI dataset \cite{thighmuscle} was used to evaluate muscle segmentation on the gracilis and sartorius muscles (MRI-Gracilis and MRI-Sartorius), with 25 randomly selected volumes.
\subsubsection{Ultrasound Segmentation}: The JNU-IFM (Intelligent Fetal Monitoring Lab of Jinan University) dataset \cite{lu2022jnu} was used to evaluate fetal head segmentation in ultrasound videos (US-Head), with 25 randomly selected sequences.

In all experiments, 5 volumes from each dataset were randomly chosen for training, and the remaining volumes were used for testing. All 2D slices (except US-Head) were extracted from axial view.

\subsection{Implementation Details}
\label{sec:implementation_details}
We followed the default settings of SAM 2 for end-to-end finetuning SLM-SAM 2, with the exception that only mask prompts were utilized, given that our study focuses on VOS tasks. All components were trainable during finetuning and all parameters were optimized jointly. The batch size was set to 1, and each input sequence was composed of 8 randomly selected consecutive 2D slices. For each segmentation task, the model was finetuned using the AdamW optimizer for 100 epochs. Due to the limited number of training samples, we evaluated using the model checkpoint from the final training epoch, in a setting similar to \cite{cai2023orthogonal,xie2024sam}. All models were trained on a single NVIDIA GeForce RTX A6000. The initial learning rates for the image encoder and the remaining model components were set to $3 \times 10^{-6}$ and $5 \times 10^{-6}$, respectively, and were updated using a cosine annealing scheduler. All 2D images were resized to a resolution of 1024 and normalized to a range of 0 to 255 at the volume level \footnote{For CT scans, intensities were clipped to the range [-1024, 2000] in Hounsfield Units (HU)}. The training data augmentation pipeline followed the SAM 2 implementation.

\subsection{Baseline Methods}
\label{sec:baseline_methods}
To evaluate the effectiveness of SLM-SAM 2, we compared its segmentation performance against state-of-the-art VOS models, including SAM 2 \cite{sam2}, Cutie \cite{cutie} and iSegFormer \cite{isegformer}. Additionally, we assessed its performance relative to leading 2D and 3D fully automated segmentation models, such as nnUNet \cite{nnunet}, Medical SAM Adapter (MSA) \cite{wu2025medical}, UNETR \cite{unetr} and Swin UNETR \cite{swinunetr}. Furthermore, we compared with ConvLSTM-based and BiConvLSTM-based methods, such as LSTM-MM-UNet \cite{lstmmmunet} and BCDUNet \cite{bcdunet}. We also compared our performance against unsupervised VOS methods, including VoxelMorph \cite{vxm} and Sli2Vol \cite{yeung2021sli2vol}. VoxelMorph and Sli2Vol are image-registration methods that learn dense correspondences between adjacent slices to propagate manual annotations across an entire volume. Cutie leverages a query-based object transformer to integrate top-down object-level memory with bottom-up pixel-level features, segmenting foreground targets through the foreground-background masked attention. Cutie+, a variant of Cutie with a larger memory bank that offers improved performance but with slower inference speed, is also included. We chose the base model for Cutie and Cutie+, initialized with pretrained weights, and followed the default training settings from official GitHub. iSegFormer integrates a 2D interactive segmentation network with a label-propagation module. In our experiments, we evaluated its propagation component with pretrained weights initialization starting from a single annotated mask. nnUNet is a robust, self-adapting framework based on the UNet architecture \cite{unet}. We evaluated the 2D nnUNet in two settings: the basic (B) and the exhaustive (E) setting. In the basic setting, the model was trained on $n$ fully annotated volumes and evaluated directly on the test set. In the exhaustive setting, we retrained a model individually for each test volume, using $n$ fully annotated volumes along with one annotated slice from that test volume. LSTM-MM-UNet uses ConvLSTM to enforce inter-slice continuity, while BCDUNet replaces the standard skip-connection in UNet with a BiConvLSTM module. We followed the official GitHub for implementation. UNETR and SwinUNETR are state-of-the-art 3D transformer-based medical image segmentation models. UNETR employs a pure transformer encoder for volumetric representation learning, while Swin UNETR utilizes swin transformer-based blocks for enhanced feature extraction. All 3D models were trained on $n$ fully annotated volumes using the AdamW optimizer with an initial learning rate of $1 \times 10^{-3}$. For fair comparison, all propagation methods propagated labels bidirectionally, used the same dataset splits and the same manual masks as the initial prompts at inference time. In addition, for automatic segmentation methods, the predictions on the selected annotated slices were replaced with the ground truth during evaluation.

\subsection{Evaluation Metrics}
\label{sec:evaluation_metrics}
We quantitatively assessed the performance of SLM-SAM 2 using the Dice Similarity Coefficient (\textbf{DSC}) and Average Symmetric Surface Distance (\textbf{ASSD}), both evaluated on 3D volumetric segmentation. Given the predicted mask $A$ and the ground truth mask $B$, the two metrics are defined as below:
\begin{equation}
    DSC(A, B) = \frac{2|A \cap B|}{|A| + |B|}
\end{equation}
DSC measures the overlap between $A$ and $B$, with values ranging from 0 and 1. Higher DSC values indicate better segmentation performance.
\begin{equation}
    ASSD(A, B) = \frac{\sum_{p \in S_A} d(p, S_B) + \sum_{q \in S_B} d(q, S_A)}{|S_A| + |S_B|}
\end{equation}
where $S_A$ and $S_B$ denote the sets of surface points of mask $A$ and $B$. $d(p, S_B)$ and $d(q, S_A)$ represent the shortest Euclidean distances from a surface point in $S_A$ and $S_B$ to the nearest point in $S_B$ and $S_A$, respectively. Compared to DSC, ASSD penalizes more on boundary errors. ASSD is non-negative and a lower ASSD indicates better segmentation performance.

For both 5-Volume and 1-Volume settings, we estimated model robustness using bootstrapping with 1,000 resamples. For the 1-Volume setting, we averaged performance from five models per sample.

\subsection{Results}
\label{sec:results}
\subsubsection{Overall Performance Comparison of SLM-SAM 2}
\begin{figure}
    \centering
    \includegraphics[width=1\linewidth]{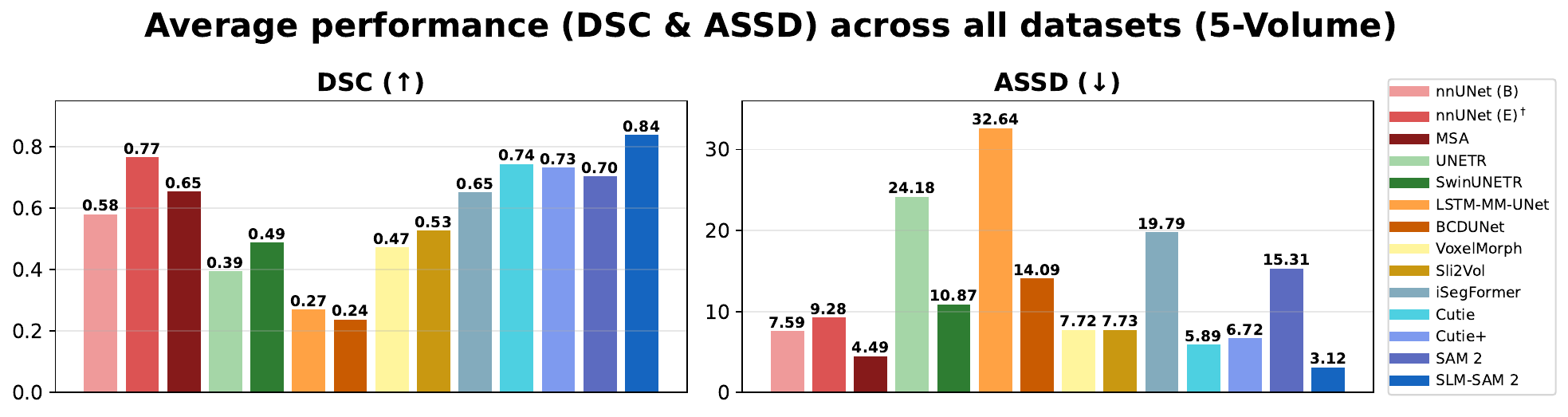}
    \caption{{\textbf{Average performance comparison across datasets (5-Volume).} SLM-SAM 2 demonstrates the best average performance compared among all baselines in both DSC and ASSD.}}
    \label{fig:ave_result}
\end{figure}

\begin{table*}[ht]
\caption{Quantitative Comparison (DSC and ASSD) for each dataset under the 5-Volume settings. Best results are in bold; runner-up is underlined. $\dag$: method requires retraining for each test volume.}

\resizebox{1\textwidth}{!}{%
\begin{tabular}{||l||cc||cc||cc||cc||cc||cc||cc||cc||cc||}
\toprule
 & \multicolumn{2}{c||}{\textbf{MRI-Kidney}} & \multicolumn{2}{c||}{\textbf{MRI-Pancreas}} & \multicolumn{2}{c||}{\textbf{CT-Pancreas}} & \multicolumn{2}{c||}{\textbf{CT-Spleen}} & \multicolumn{2}{c||}{\textbf{CT-Heart}} & \multicolumn{2}{c||}{\textbf{CT-Femur}} & \multicolumn{2}{c||}{\textbf{MRI-Gracilis}} & \multicolumn{2}{c||}{\textbf{MRI-Sartorius}} & \multicolumn{2}{c||}{\textbf{US-Head}}  \\
\midrule
 \textbf{Methods} & DSC ($\uparrow$) & ASSD ($\downarrow$) & DSC ($\uparrow$) & ASSD ($\downarrow$) & DSC ($\uparrow$) & ASSD ($\downarrow$) & DSC ($\uparrow$) & ASSD ($\downarrow$) & DSC ($\uparrow$) & ASSD ($\downarrow$) & DSC ($\uparrow$) & ASSD ($\downarrow$) & DSC ($\uparrow$) & ASSD ($\downarrow$) & DSC ($\uparrow$) & ASSD ($\downarrow$) & DSC ($\uparrow$) & ASSD ($\downarrow$) \\
\midrule
\multicolumn{19}{||c||}{\textbf{I. 2D Models}} \\
\midrule
nnUNet (B) & 0.6497 & 5.8917 & 0.3348 & 8.1976 & 0.1445 & 22.6483 & 0.2731 & 10.8427 & 0.6426 & 7.5425 & 0.9013 & 1.8705 & 0.7083 & 1.9009 & 0.6818 & 1.4850 & 0.6991 & 10.8537 \\
nnUNet (E)$^{\dag}$ & \underline{0.9073} & 2.8406 & \underline{0.6234} & 8.8195 & 0.4337 & 23.2610 & 0.7548 & 29.3823 & \textbf{0.8013} & 9.1686 & \underline{0.9476} & \underline{1.5655} & 0.7843 & 1.7841 & \underline{0.7562} & \underline{1.4003} & 0.7817 & 9.2342 \\
MSA & 0.8366 & 1.7724 & 0.5203 & \textbf{4.8257} & 0.1514 & 10.6728 & 0.6367 & 3.4611 & 0.7020 & \textbf{5.4726} & 0.9223 & 3.3323 & 0.6345 & 1.6441 & 0.5967 & 2.4229 & 0.7356 & 6.9148 \\
\midrule
\multicolumn{19}{||c||}{\textbf{II. 3D Models}} \\
\midrule
UNETR & 0.4661 & 22.5008 & 0.2754 & 13.3468 & 0.0320 & 76.1877 & 0.1548 & 51.7418 & 0.5727 & 21.5100 & 0.5524 & 10.3962 & 0.4691 & 3.8774 & 0.3260 & 7.7069 & 0.5376 & 18.6205 \\
SwinUNETR & 0.6508 & 4.6814 & 0.3778 & 8.8737 & 0.1525 & 13.4887 & 0.3182 & 21.0385 & 0.5937 & 17.1132 & 0.5948 & 8.6549 & 0.5324 & 3.9244 & 0.4981 & 3.7137 & 0.5796 & 14.2872 \\
\midrule
\multicolumn{19}{||c||}{\textbf{III. ConvLSTM/BiConvLSTM-based Models}} \\
\midrule
LSTM-MM-UNet & 0.4483 & 18.7278 & 0.0741 & 18.4087 & 0.1178 & 21.6110 & 0.3521 & 14.5100 & 0.1877 & 46.1875 & 0.4406 & 31.9848 & 0.1423 & 54.3272 & 0.0935 & 51.4670 & 0.5311 & 30.1221 \\
BCDUNet & 0.4253 & 4.4763 & 0.0651 & 21.7007 & 0.1100 & 26.6618 & 0.1191 & 7.2795 & 0.4579 & 25.5236 & 0.0539 & 15.8434 & 0.2282 & 4.4915 & 0.2674 & 3.3482 & 0.4005 & 10.8755 \\
\midrule
\multicolumn{19}{||c||}{\textbf{IV. Unsupervised VOS Models}} \\
\midrule
VoxelMorph & 0.3854 & 5.8693 & 0.2501 & 13.7858 & 0.4307 & 15.6178 & 0.7806 & 2.6862 & 0.5211 & 6.6286 & 0.3411 & 8.3026 & 0.5707 & 3.0872 & 0.5163 & 2.8654 & 0.4984 & 10.8782 \\
Sli2Vol & 0.8287 & \underline{1.6272} & 0.4676 & 10.9631 & 0.4636 & 13.0736 & 0.7549 & 5.3969 & 0.6243 & 6.3413 & 0.3042 & 9.1010 & 0.5214 & 3.7066 & 0.4352 & 4.0077 & 0.4128 & 15.3546 \\
\midrule
\multicolumn{19}{||c||}{\textbf{V. Semi-supervised VOS Models}} \\
\midrule
iSegFormer & 0.8067 & 8.9592 & 0.4358 & 17.7530 & 0.5016 & 14.4014 & 0.8605 & 7.8476 & 0.6496 & 13.9058 & 0.5769 & 47.0773 & \underline{0.8156} & 1.7567 & 0.7347 & 4.5420 & 0.5239 & 51.1868 \\
Cutie & 0.8976 & 1.9001 & 0.5340 & 6.6390 & \underline{0.6806} & 8.3856 & \underline{0.9629} & 0.5550 & \underline{0.7855} & \underline{6.2811} & 0.6303 & 16.7382 & 0.7081 & 2.1894 & 0.6813 & 1.5448 & 0.8500 & 3.1714 \\
Cutie+ & 0.8997 & 1.6683 & 0.5572 & \underline{6.3690} & 0.6628 & \underline{7.7925} & 0.9574 & 0.5197 & 0.7031 & 12.2835 & 0.6285 & 17.0783 & 0.7209 & 2.1434 & 0.6839 & 1.5357 & 0.8501 & 3.1133 \\
SAM 2 & 0.8668 & 5.9103 & 0.2341 & 30.4152 & 0.6381 & 9.1385 & \textbf{0.9633} & \textbf{0.4508} & 0.6925 & 15.6042 & 0.6267 & 52.3565 & 0.7926 & \underline{1.4828} & 0.7045 & 1.4530 & \underline{0.8635} & \textbf{2.3445} \\
\rowcolor{gray!30}
SLM-SAM 2 & \textbf{0.9368} & \textbf{1.3643} & \textbf{0.6294} & 8.4185 & \textbf{0.7255} & \textbf{4.0753} & 0.9607 & \underline{0.4767} & 0.7655 & 8.2143 & \textbf{0.9643} & \textbf{0.4298} & \textbf{0.8543} & \textbf{0.7276} & \textbf{0.8118} & \textbf{0.7524} & \textbf{0.8764} & \underline{2.4207} \\
\bottomrule
\end{tabular}
}
\label{tab:evaluation_results}
\end{table*}

\begin{figure*}[h]
    \centering
    \includegraphics[width=1\linewidth]{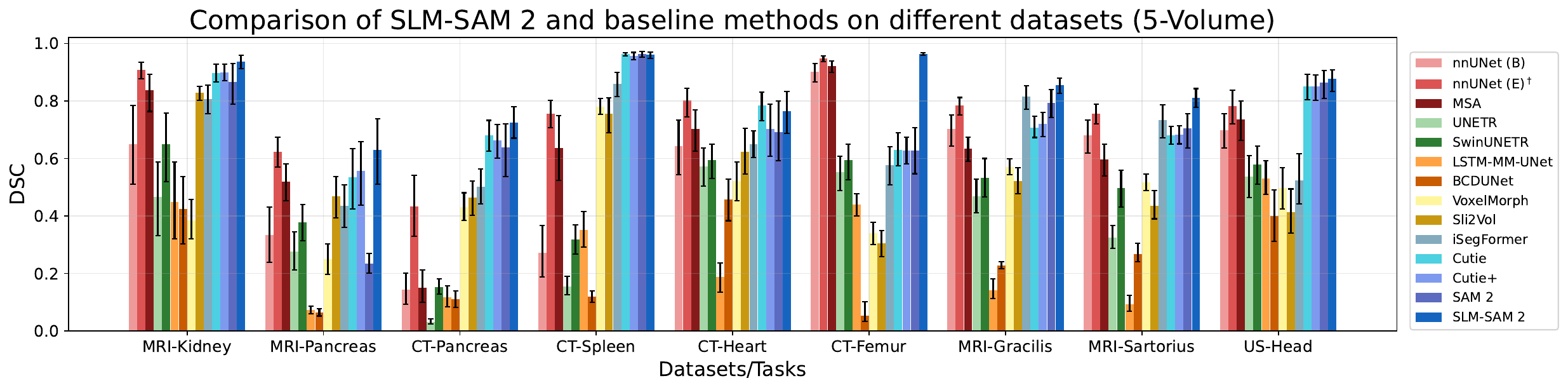}
    \caption{\textbf{Performance results (DSC) of each method across datasets (5-Volume).} SLM-SAM 2 outperforms all baselines on most datasets by a significant margin. Confidence intervals are estimated using bootstrapping with 1000 resamples.}
    \label{fig:all_result}
\end{figure*}

The average performance of SLM-SAM 2 with other baselines is presented in Fig. \ref{fig:ave_result}. Overall, we observed that SLM-SAM 2 consistently outperforms the competing methods by a significant margin, achieving improvements of over 0.07 in DSC and 2.77 in ASSD at least. In particular, while SAM 2 achieves a DSC of 0.70, SLM-SAM 2 attains a DSC of 0.84, thereby demonstrating the benefits of incorporating both short-term and long-term memory banks in mask propagation.

VOS methods generally show superior performance in slice propagation tasks while maintaining comparable computational costs. One explanation is iSegFormer, Cutie and SAM 2 are initialized with pretrained weights, leading to outperformance when finetuned on small datasets. Notably, Cutie and Cutie+ slightly outperformed SAM 2, achieving DSC of 0.74 and 0.73, respectively. This is reasonable as Cutie is found to be one of the most competitive methods relative to SAM 2 \cite{sam2}.

The runner-up method, nnUNet (E), requires retraining for each new volume, which leads to a considerable increase in time and computational resources. However, excluding the annotated slice in the training (i.e., nnUNet (B)) results in a significant performance degradation (0.19 reduction in DSC). Moreover, MSA, which leverages the SAM's pretrained encoder, exhibits superior performance than nnUNet (B).

The 3D models, UNETR and SwinUNETR, show suboptimal performance with DSC values of 0.39 and 0.49, respectively. One reason could be they do not leverage annotated slice information during either the training or inference stages.

Moreover, unsupervised methods, VoxelMorph and Sli2Vol, exhibit lower DSC performance compared to the VOS-based methods. This underperformance can be attributed to the training on unlabeled data, which focuses on learning general information transformation between slices and may not directly learn task-specific features for accurate segmentation. Additionally, ConvLSTM-based methods, LSTM-MM-UNet and BCDUNet, built on UNet backbones without pretrained encoders, yield the lowest segmentation accuracies compared to other evaluated approaches.

\begin{figure*}[!h]
    \centering
    \includegraphics[width=1\linewidth]{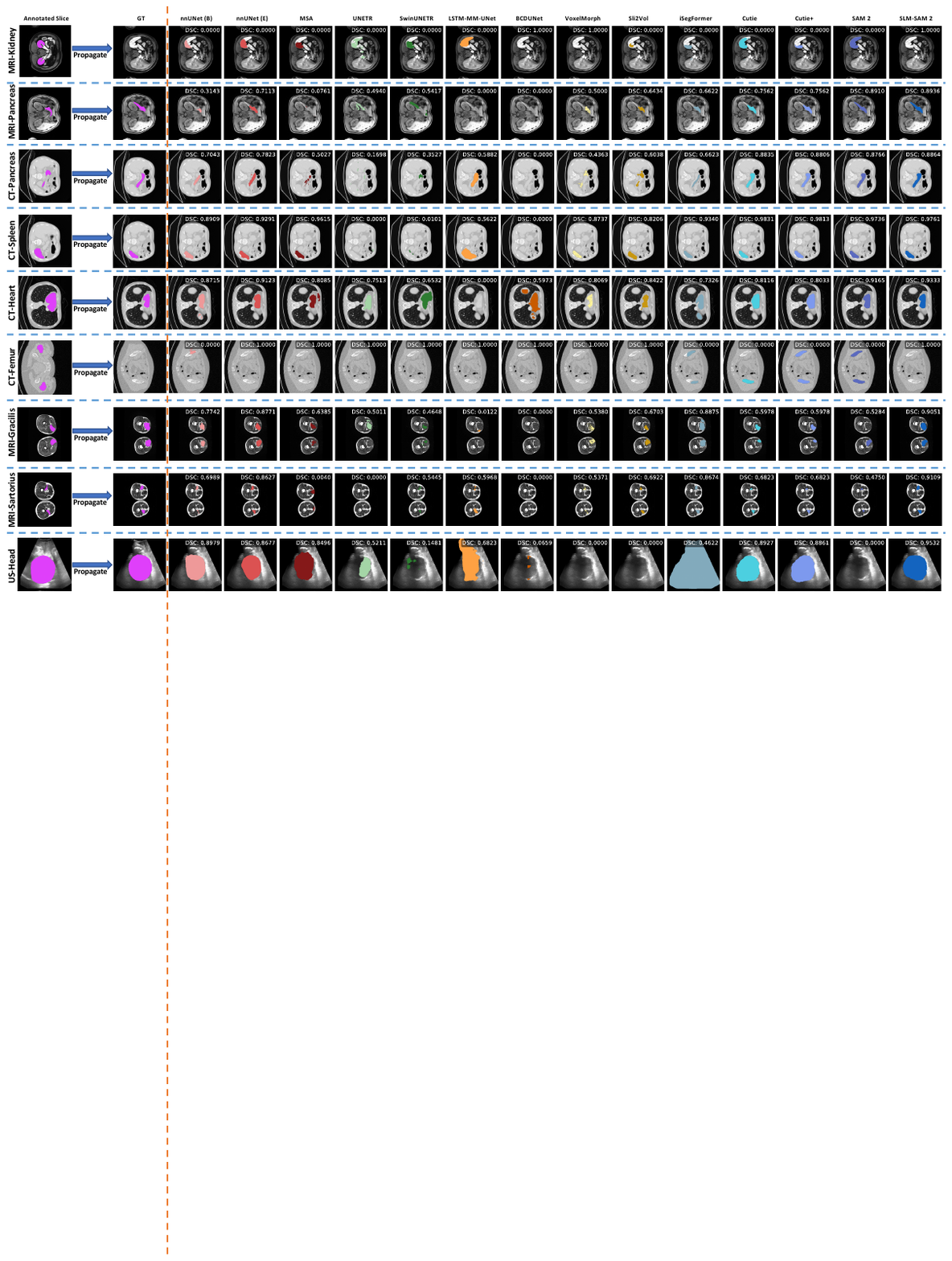}
    \caption{\textbf{Qualitative comparison on each dataset (5-Volume).} The leftmost column shows the annotated slices, followed by the ground truth (GT) and segmentations from each method. For each dataset, we present results on both target-present and target-absent slices.}
    \label{fig:all_result_visual}
\end{figure*}

\subsubsection{Performance Comparison of Each Dataset/Task}
The quantitative results (DSC and ASSD) of SLM-SAM 2 and all competing methods across the datasets/tasks are presented in Table \ref{tab:evaluation_results} and Fig. \ref{fig:all_result}. Overall, SLM-SAM 2 outperforms the other baselines in terms of DSC on most datasets. For the CT-Spleen tasks, all VOS methods achieve a DSC of around 0.96, performing comparable to each other. On CT-Heart, SLM-SAM 2 (DSC: 0.7655) underperforms nnUNet (E) (DSC: 0.8013) and slightly underperforms Cutie (DSC: 0.7855). However, nnUNet (E) requires retraining for each new volume, limiting its scalability. It is notably that SLM-SAM 2 consistently outperforms SAM 2 across organs, bones, muscles, and fetal head, demonstrating its improved robustness in mask propagation tasks. Moreover, among 2D method, nnUNet (E) consistently outperform nnUNet (B) for all datasets. This is expected as nnUNet (E) leverages an additional annotated slice from each test volume. For the 3D methods, SwinUNETR demonstrates superior performance compared to UNETR on all datasets. For unsupervised VOS methods, VoxelMorph outperforms Sli2Vol on certain datasets, such as CT-Femur, MRI muscle tasks and US-Head while Sli2Vol achieves better performance than VoxelMorph on both MRI and CT organ tasks. For ConvLSTM-based methods, BCDUNet outperforms LSTM-MM-UNet on CT-Heart and MRI muscle tasks, while LSTM-MM-UNet achieves superior performance on the others.

\begin{figure*}[h]
    \centering
    \includegraphics[width=1\linewidth]{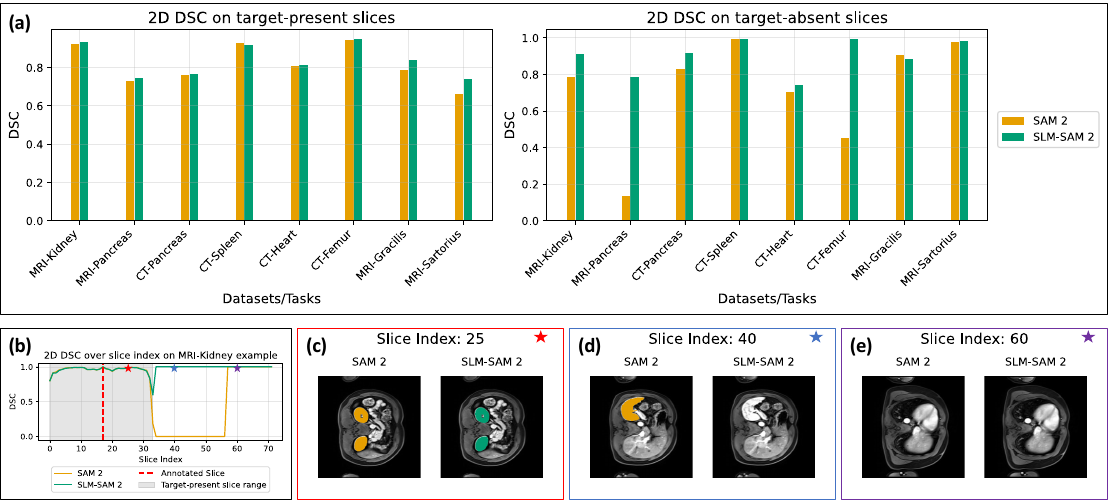}
    \caption{\textbf{Performance results on target-present and target-absent slices (5-Volume).} (a) presents 2D slice-level DSC for both target-present slices and target-absent slices across datasets. (b) plots 2D slice-level DSC over slice index on an MRI-Kidney example. (c)-(e) provide segmentation comparison between SAM 2 and SLM-SAM 2 on representative slices from the same MRI-Kidney volume.}
    \label{fig:target_vs_empty}
\end{figure*}

Fig. \ref{fig:all_result_visual} presents the qualitative results of all methods across all datasets. For each dataset, the annotated slices from the selected volume are shown in the leftmost column. The remaining columns display the ground truth and the propagated masks on representative slices. For target-present slices (i.e., slices where target is present), it can be observed that SLM-SAM 2 can achieve a better performance in segmenting small and anatomically challenging structures, such as the MRI-Gracilis and MRI-Sartoruis muscles (as shown in Fig. \ref{fig:target_vs_empty} (a)), where accurate delineation is often hindered by low contrast and ambiguous boundaries. For larger and more distinct anatomical structures, including CT-Spleen, CT-Heart and CT-Femur, the segmentation performance of SLM-SAM 2 seems to be comparable to that of some other baselines. 

\begin{figure*}[h]
    \centering
    \includegraphics[width=1\linewidth]{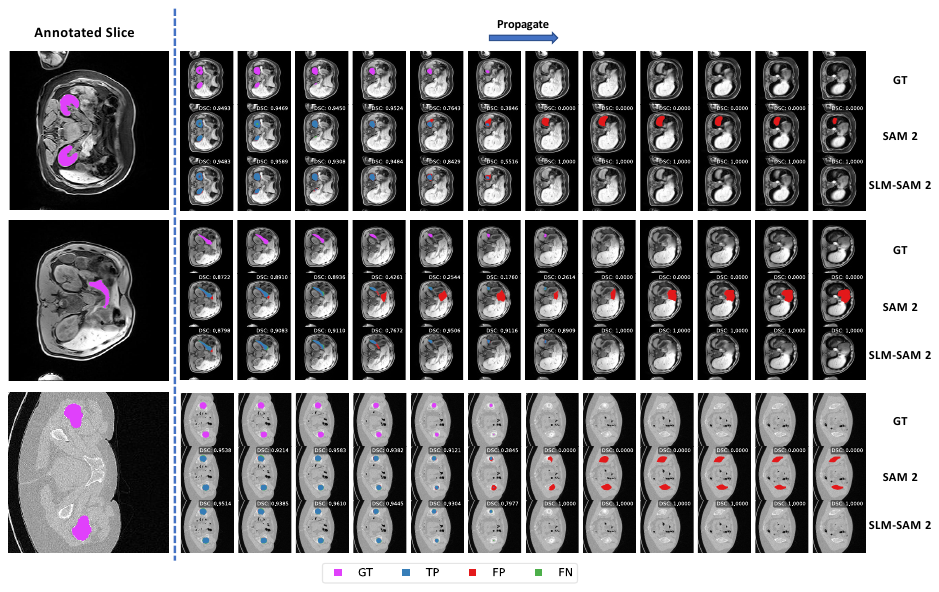}
    \caption{\textbf{Qualitative comparison of propagation between SLM-SAM 2 and SAM 2.} SLM-SAM 2 performs robust at object boundaries, effectively alleviating over-propagation issue. True positive, false positive and false negative predictions are depicted in blue, red and green.}
    \label{fig:propagation}
\end{figure*}

For target-absent slices (i.e., slices where the target is absent), SAM 2 and several other VOS methods tend to over-propagate masks of segmentation into regions where the target is no longer present, particularly in adjacent anatomical structures with similar intensity. For instance, in MRI-Kidney, iSegFormer, Cutie, Cutie+ and SAM 2 sometimes over-propagate the kidney masks onto the spleen. Similarly, in CT-Femur, other VOS methods may segment the hip bones as the femurs, likely due to their similar appearance and spatial overlap in axial 2D slices. These challenges are not limited to VOS methods. Competing 2D and 3D segmentation models, such as nnUNet (E), UNETR and SwinUNETR, can also mis-segment adjacent anatomical structures. For example, these models may incorrectly segment the spleen in MRI-Kidney tasks. However, experimental results demonstrate that SLM-SAM 2, which incorporates dynamically learned short-term and long-term memory mechanisms, can effectively alleviate this issue (Fig. \ref{fig:all_result_visual}, Fig. \ref{fig:target_vs_empty} (a) and Fig. \ref{fig:propagation}), leading to more accurate and robust models for slice propagation.

\subsubsection{SLM-SAM 2 v.s. SAM 2 on Over-propagation} We further investigated the benefits of integrating both short-term and long-term memory in SLM-SAM 2, compared to SAM 2. To do this, we computed the 2D slice-wise DSC on target-present slices and target-absent slices, as shown in Fig. \ref{fig:target_vs_empty} (a). 

\begin{figure*}[h]
    \centering
    \includegraphics[width=1\linewidth]{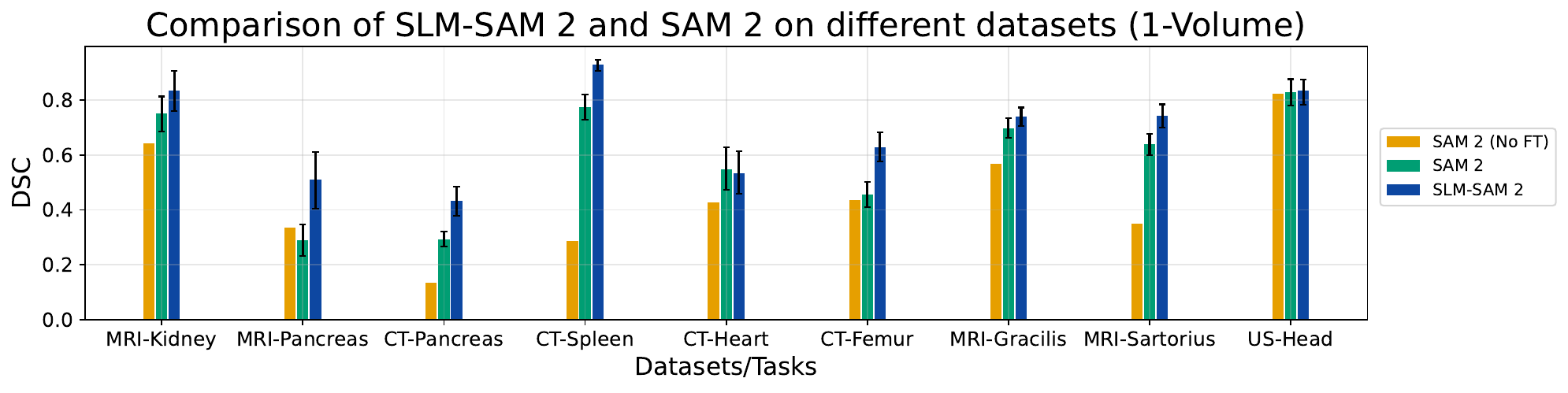}
    \caption{\textbf{Performance results (DSC) of SAM 2 and SLM-SAM 2 (1-Volume).} SLM-SAM 2 outperforms SAM 2 on most datasets under the 1-Volume settings, excluding CT-Heart, where both methods perform similarly. 5 volumes are randomly selected for each dataset.}
    \label{fig:all_result_1_shot}
\end{figure*}

\begin{figure}[h]
    \centering
    \includegraphics[width=1\linewidth]{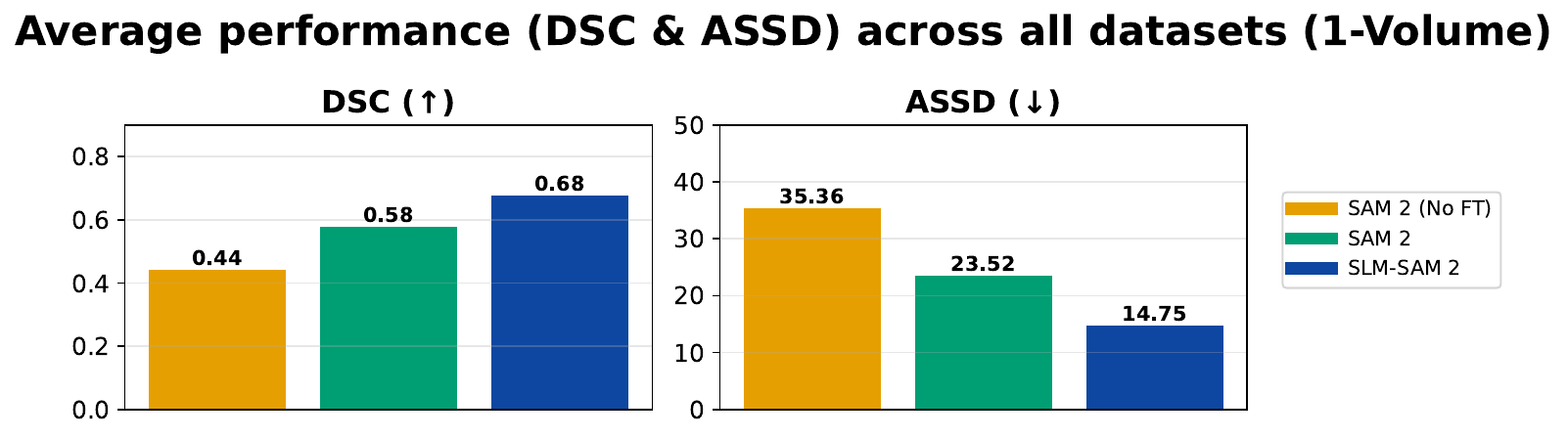}
    \caption{\textbf{Average performance comparison across dataset (1-Volume).} SLM-SAM 2 shows the better average performance compared with SAM 2 under the 1-Volume setting.}
    \label{fig:ave_result_1_shot}
\end{figure}

For most datasets except two MRI muscles datasets, SLM-SAM 2 and SAM 2 achieve comparable performance on target-present slices. However, for target-absent slices, SLM-SAM 2 demonstrates significant outperformance compared to SAM 2, with particularly notable improvements observed on MRI-Kidney, MRI-Pancreas, CT-Pancreas and CT-Femur. In addition, SLM-SAM 2 exhibits superior performance on both target-present and target-absent slices on US-Head.

Fig. \ref{fig:target_vs_empty} (b)-(e) presents more direct and illustrative results. Specifically, Fig. \ref{fig:target_vs_empty} (b) depicts the 2D DSC over the slice index for an example from MRI-Kidney dataset. Fig. \ref{fig:target_vs_empty} (c), (d) and (e) provide a qualitative comparison of the segmentation performance between SLM-SAM 2 versus SAM 2 on the selected representative slices in the MRI-Kidney example. Firstly, it can be seen that the 2D DSC on target-present slices is similar between SLM-SAM 2 and SAM 2, as shown in Fig. \ref{fig:target_vs_empty} (c). However, once the kidney is no longer present (i.e., beyond the gray panel range), the 2D DSC of SAM 2 drops to zero, whereas SLM-SAM 2 maintains a 2D DSC of 1. As illustrated in Fig. \ref{fig:target_vs_empty} (d), this is because SAM 2 over-propagates the kidney mask into the spleen region. A potential explanation is the similar intensity between the spleen and kidney, which may obscure their boundaries. Furthermore, as depicted in Fig. \ref{fig:target_vs_empty} (e), SAM 2 stops propagating the mask after the spleen disappears from the images. This may be because the adjacent lung tissue has a much lower intensity (i.e., darker) in MRI, making the boundary more distinct.

On target-present slices, SLM-SAM 2 achieves an overall better performance compared to SAM 2 in MRI muscle datasets and US-Head, while maintaining comparable performance across other tasks. By incorporating a dynamic learning mechanism that integrates both short-term and long-term memory, SLM-SAM 2 exhibits improved robustness at target boundaries, significantly mitigating over-propagation problem. We argue that such vague boundary situations are common in medical imaging, like unclear boundaries often occur between the femur and hip bones in CT scans, as shown in Fig. \ref{fig:all_result_visual}. These challenges may result from factors such as intensity similarities between the target and neighboring structures, close proximity or direct contact between anatomical structures, and limitations in scan resolution (i.e., slice thickness or spacing between slices). Therefore, the ability of SLM-SAM 2 to better handle boundary ambiguity makes it a promising approach for medical image segmentation.

\begin{table*}[ht]
\caption{Quantitative results (DSC) of ablation study under the 5-Volume settings. Best results are in bold; runner-up is underlined.}

\resizebox{1\textwidth}{!}{%
\begin{tabular}{||l|c|c||c|c|c|c|c|c|c|c|c||}
\toprule
& & & \textbf{MRI-Kidney} & \textbf{MRI-Pancreas} & \textbf{CT-Pancreas} & \textbf{CT-Spleen} & \textbf{CT-Heart} & \textbf{CT-Femur} & \textbf{MRI-Gracilis} & \textbf{MRI-Sartorius} & \textbf{US-Head} \\
\midrule
 \textbf{Methods} & $\mathcal{M}_1$ & $\mathcal{M}_2$ & DSC ($\uparrow$) & DSC ($\uparrow$) & DSC ($\uparrow$) & DSC ($\uparrow$) & DSC ($\uparrow$) & DSC ($\uparrow$) & DSC ($\uparrow$) & DSC ($\uparrow$) & DSC ($\uparrow$) \\
\midrule
$\mathcal{M}_{O}$ (No FT) & $\mathcal{M}_{O}$ & \xmark & 0.6428 & 0.3358  & 0.1345 & 0.2858 & 0.4271 & 0.4359 & 0.5662 & 0.3483 & 0.8233 \\
$\mathcal{M}_{O}$ & $\mathcal{M}_{O}$ & \xmark & 0.8668 & 0.2341 & 0.6381 & \textbf{0.9633} & 0.6925 & 0.6267 & \underline{0.7926} & 0.7045 & \underline{0.8635} \\
$\mathcal{M}_{R}=7$ & $\mathcal{M}_{R}=7$ & \xmark & \underline{0.9088} & 0.3009 & 0.6552 & 0.9590 & 0.6650 & 0.9232 & 0.7043 & 0.4989 & 0.8270 \\
$\mathcal{M}_{R}=1$ & $\mathcal{M}_{R}=1$ & \xmark & 0.9018 & \underline{0.4866} & 0.6637 & 0.9494 & 0.7138 & 0.9351 & 0.7900 & \underline{0.7882} & 0.8261 \\
\midrule
$\mathcal{M}_{O}+(\mathcal{M}_{R}=7)$ & $\mathcal{M}_{O}$ & $\mathcal{M}_{R}=7$ & 0.8916 & 0.3765 & \underline{0.6690} & \underline{0.9621} & \underline{0.7177} & \underline{0.9559} & 0.7765 & 0.6836 & 0.8538 \\
$\mathcal{M}_{O}+(\mathcal{M}_{R}=1)$ & $\mathcal{M}_{O}$ & $\mathcal{M}_{R}=1$ & \textbf{0.9368} & \textbf{0.6294} & \textbf{0.7255} & 0.9607 & \textbf{0.7655} & \textbf{0.9643} & \textbf{0.8543} & \textbf{0.8118} & \textbf{0.8764} \\
\bottomrule
\end{tabular}
}
\label{tab:ablation_results}
\end{table*}

\subsubsection{Experiments on 1-Volume Setting}

In practical scenarios, practitioners may be interested in annotating a single volume and applying the finetuned model on that volume directly to annotate new data to save time. Thus, we also investigated the performance of SLM-SAM 2 under the 1-Volume setting in comparison with SAM 2. Specifically, we randomly selected five different volumes for training and report the average performance of the five models for each dataset. We followed the same training and inference settings as in the 5-Volume experiments, with the same test sets. Fig. \ref{fig:ave_result_1_shot} shows that SLM-SAM 2 achieves better average performance (DSC: 0.68, ASSD: 14.75) than SAM 2 (DSC: 0.58, ASSD: 23.52), demonstrating that SLM-SAM 2 also provides a more accurate and robust models under the 1-Volume setting. Furthermore, Fig. \ref{fig:all_result_1_shot} presents the performance on each dataset, where SLM-SAM 2 consistently outperforms SAM 2 by a significant margin on most datasets.

\subsubsection{Ablation Study}

In this section, we present an ablation study to investigate the impact of different memory bank configurations on model performance. Specifically, we examined how varying the selection and combination of slices within the memory bank influences the results. We denote the baseline SAM 2 configuration as $\mathcal{M}_{O}$, which consists of an original memory bank containing one annotated slice and six additional recent slices. Our SLM-SAM 2 is represented as $\mathcal{M}_{O} + (\mathcal{M}_{R}=1)$, which dynamically integrates the original SAM 2 memory bank ($\mathcal{M}_{O}$) with a recent memory bank that includes the most recent slice. Furthermore, we evaluated the performance when using only the recent memory bank, considering two configurations: one with the seven most recent slices ($\mathcal{M}_{R}=7$) and another with only the single most recent slice ($\mathcal{M}_{R}=1$). In addition, we explored the combination of the original memory bank $\mathcal{M}_{O}$ with the recent bank containing the seven most recent slices, denoted as $\mathcal{M}_{O} + (\mathcal{M}_{R}=7)$.

\begin{figure}[!h]
    \centering
    \includegraphics[width=1\linewidth]{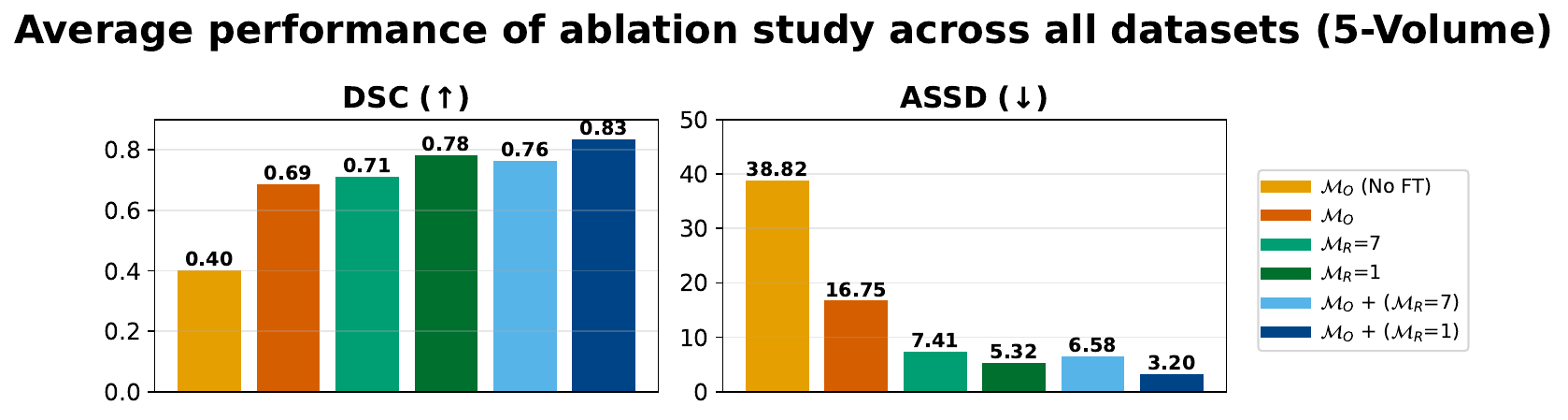}
    \caption{\textbf{Average performance comparison of ablation study.} SLM-SAM 2 (i.e., $\mathcal{M}_{O} + (\mathcal{M}_{R}=1)$) achieves best results compared with other memory bank configurations in both DSC and ASSD.}
    \label{fig:ave_result_ablation}
\end{figure}

\begin{figure}[!h]
    \centering
    \includegraphics[width=1\linewidth]{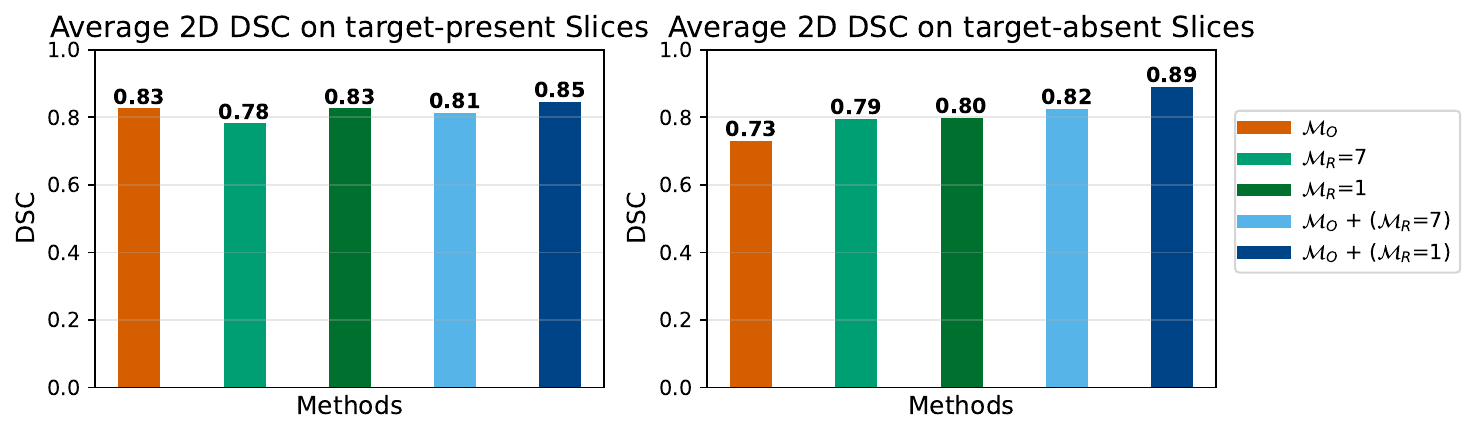}
    \caption{\textbf{Average performance results of ablation study on target-present and target-absent slices.} SLM-SAM 2 achieves best performance on both target-present and target-absent slices in terms of DSC.}
    \label{fig:ave_result_ablation_target_empty}
\end{figure}

\begin{figure*}[h]
    \centering
    \includegraphics[width=1\linewidth]{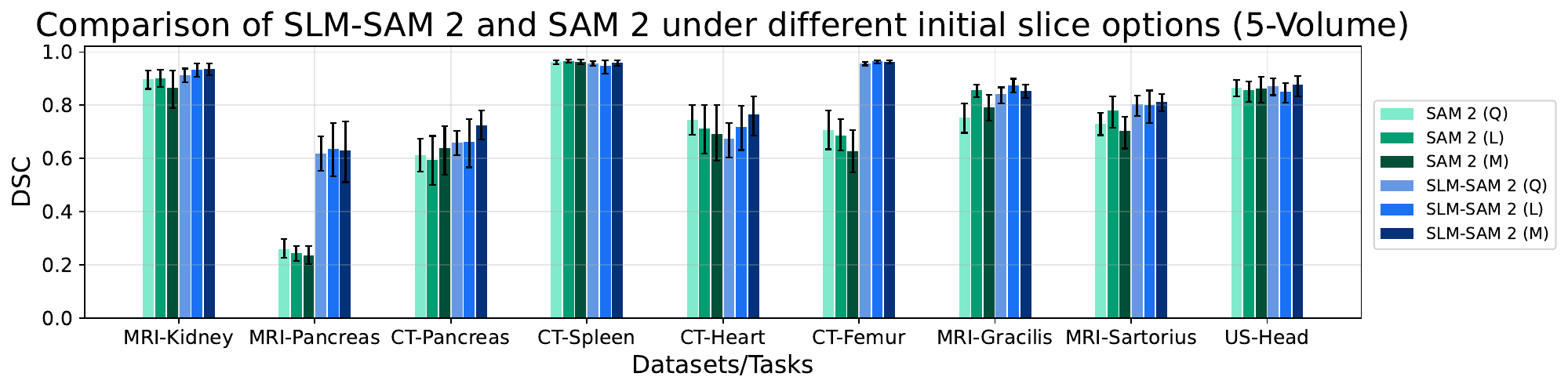}
    \caption{\textbf{Performance results (DSC) of SAM 2 and SLM-SAM 2 under different initial slice options (5-Volume).} SLM-SAM 2 outperforms SAM 2 on most datasets across different slice initializations, while performance within each method remain comparable under varying initial slice settings.}
    \label{fig:all_result_initial_slice}
\end{figure*}

As shown in Fig. \ref{fig:ave_result_ablation}, it can be seen that using only recent memory (i.e., $\mathcal{M}_{R}=7$ and $\mathcal{M}_{R}=1$) significantly outperforms $\mathcal{M}_{O}$. Specifically, Fig. \ref{fig:ave_result_ablation_target_empty} shows that utilizing recent memory alone achieves superior performance on target-absent slices, with $\mathcal{M}_{R}=7$ and $\mathcal{M}_{R}=1$ achieving DSC of 0.79 and 0.80, respectively, compared to 0.73 for $\mathcal{M}_{O}$. A potential explanation is that retaining annotated slices in the single memory bank (e.g., $\mathcal{M}_{O}$) may lead to less robustness at object boundaries and an increased risk of over-propagation errors.

\begin{figure*}[h]
    \centering
    \includegraphics[width=1\linewidth]{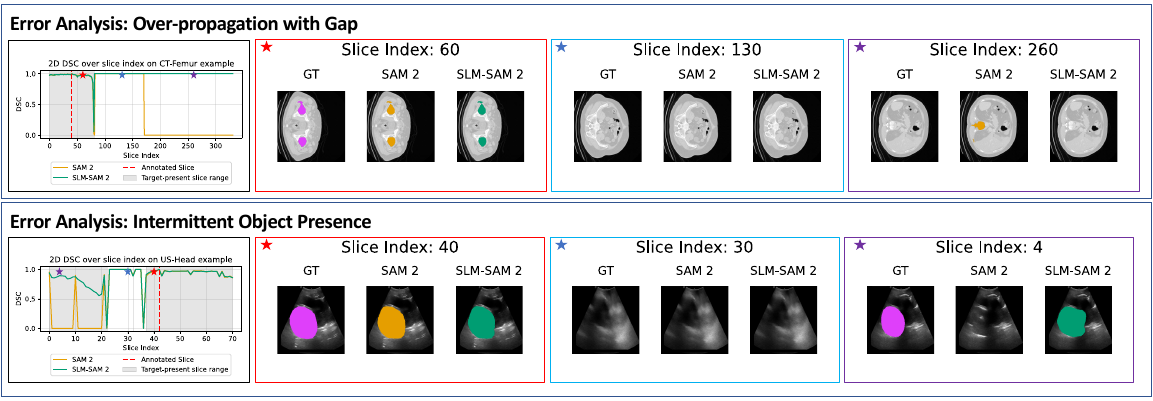}
    \caption{\textbf{Qualitative results of error analysis in VOS.} SLM-SAM 2 exhibits greater robustness than SAM 2 in handling over-propagation with gaps and intermittent object reappearances.}
    \label{fig:error_analysis}
\end{figure*}

Moreover, we find that combining the original memory bank $\mathcal{M}_{O}$ with recent memory banks, such as $\mathcal{M}_{O} + (\mathcal{M}_{R}=1)$ and $\mathcal{M}_{O} + (\mathcal{M}_{R}=7)$, results in improved performance compared to using either $\mathcal{M}_{O}$ or the respective recent memory bank alone (shown in Fig. \ref{fig:ave_result_ablation}). This demonstrates the effectiveness of the proposed dynamically learned memory mechanism, which integrates information from multiple memory banks.

Notably, as shown in Table \ref{tab:ablation_results}, $\mathcal{M}_{O} + (\mathcal{M}_{R}=1)$ consistently outperforms $\mathcal{M}_{O} + (\mathcal{M}_{R}=7)$, with the exception of CT-Spleen, where both configurations achieve a comparable DSC of approximately 0.96. This finding aligns with our expectations, as combining short-term memory bank ($\mathcal{M}_{R}=1$) and $\mathcal{M}_{O}$ enables the model to leverage both gradual temporal information and direct attention to the most recent slice.

Overall, $\mathcal{M}_{O} + (\mathcal{M}_{R}=1)$ achieves the best performance on both target-present and target-absent slices, highlighting the advanced memory mechanism design of SLM-SAM 2.

In addition, we observe that finetuned SAM 2 underperforms non-finetuned SAM 2 on the MRI-Pancreas dataset. This may be attributed to overfitting, a common challenge in few-shot finetuning \cite{sun2022singular,kumar2022fine}.

\subsection{Initial Slice Sensitivity Analysis}
\label{sec:initial_slice_sensitivity}
In this project, we primarily studied label propagation from the middle slice of the object (M). However, since the middle slice may not always be optimal, we also assessed performance sensitivity to initial slice choice by adding experiments starting propagation from (1) L: the slice with the largest object area, and (2) Q: the quarter slices of the object. For Q, we selected both the first-quarter and third-quarter slices and compute the average performance to mitigate ordering effects.
\begin{table}[ht]
\tiny
\centering
\caption{Average performance (DSC) of SAM 2 and SLM-SAM 2 across all datasets under different initial slice selection options. Best results are in bold.}

\resizebox{0.35\textwidth}{!}{%
\begin{tabular}{c|c|c|c}
\toprule
\textbf{Methods} & \textbf{Q} & \textbf{L} & \textbf{M} \\
\midrule
 SAM 2 & 0.7261 & 0.7296 & 0.7042 \\
 SLM-SAM 2 & \textbf{0.8107} & \textbf{0.8231} & \textbf{0.8390} \\
\bottomrule
\end{tabular}
}
\label{tab:initial_slices}
\end{table}

As shown in Table \ref{tab:initial_slices} and Fig. \ref{fig:all_result_initial_slice}, SLM-SAM 2 consistently outperforms SAM 2 across all three initial slice options both on average and across most datasets, demonstrating the robustness of the dynamical short-long memory module regardless of initial points. Moreover, the performance within each method is comparable across different initial slice selections.

\subsection{Propagation Error Analysis}
\label{sec:propagation_error_analysis}
In addition to over-propagation, we also evaluated SLM-SAM 2 under other common error scenarios in propagation, comparing its performance to SAM 2.

\subsubsection{Over-Propagation with Gap}
\label{sec:overpropagation_with_gap}
Over-propagation can occur immediately after the ground truth ends or after a gap. Fig. \ref{fig:target_vs_empty} (b)-(e)
and \ref{fig:propagation} have demonstrated SLM-SAM 2 handles immediate over-propagation more effectively than SAM 2. As shown in Fig. \ref{fig:error_analysis}, in cases where SAM 2 experiences over-propagation after a gap, SLM-SAM 2 can also address this issue.

\subsubsection{Intermittent Object Presence}
Another challenging scenarios in VOS tasks is intermittent object presence, where the target disappears and then reappears. In medical imaging, this occurs more frequently in ultrasound and surgical videos. We qualitatively evaluated such cases in Fig. \ref{fig:error_analysis}, where the fetal head vanishes for several slices and then reappears. It can be observed that SAM 2 struggles to recover the segmentation, while SLM-SAM 2 can handle the interruption more robustly.

\subsubsection{Premature Termination}
In VOS, labels may terminate prematurely, resulting in missing predicted masks at object boundaries. To quantify this error, we computed the \textbf{FNSR} (false negative slice ratio), and the \textbf{PFNSR} (premature false negative slice ratio) as follows:

\begin{equation}
    FNSR = \frac{\text{\# of false negative slices}}{\text{\# of target-present slices}}
\end{equation}

\begin{equation}
    PFNSR = \frac{\text{\# of premature false negative slices}}{\text{\# of target-present slices}}
\end{equation}

\begin{table}[ht]
\tiny
\centering
\caption{Comparison of SLM-SAM 2 and SAM 2 in terms of FNSR and PFNSR.}

\resizebox{0.35\textwidth}{!}{%
\begin{tabular}{c|c|c}
\toprule
\textbf{Methods} & \textbf{FNSR} & \textbf{PFNSR} \\
\midrule
 SAM 2 & 3.205 $\%$ & 2.669 $\%$ \\
 SLM-SAM 2 & 2.244 $\%$ & 2.111 $\%$ \\
\bottomrule
\end{tabular}
}
\label{tab:premature_termination}
\end{table}

where false negative slices are target-present slices with empty predictions. Premature false negative slices are the consecutive false negative slices at the boundaries of the target-present range (i.e., from the first false negative slice up to the first target-present slice with any predicted mask, and from the last false negative slice backward to the last target-present slice with any predicted mask). Thus, PFNSR quantifies the fraction of target-present slices missed due to premature termination.

As shown in Table. \ref{tab:premature_termination}, SLM-SAM 2 yields fewer false negative slices than SAM 2. Moreover, SLM-SAM 2 achieves a lower PFNSR, demonstrating superior robustness to premature termination at object boundaries.

\subsection{Multi-label Propagation}
We evaluated the ability of SLM-SAM 2 on handling multi-label propagation under three settings: (1) I+I: models trained and tested separately on each class, (2) M+I: a single model trained on combined multi-class labels but evaluated per class, and (3) M+M: a single model trained and evaluated simultaneously on all classes by annotating slices containing every target. We conducted the experiments on the MRI-Gracilis and MRI-Sartorius datasets, where each slice contains labels of gracilis and sartorius muscles.

\begin{figure}[!h]
    \centering
    \includegraphics[width=1\linewidth]{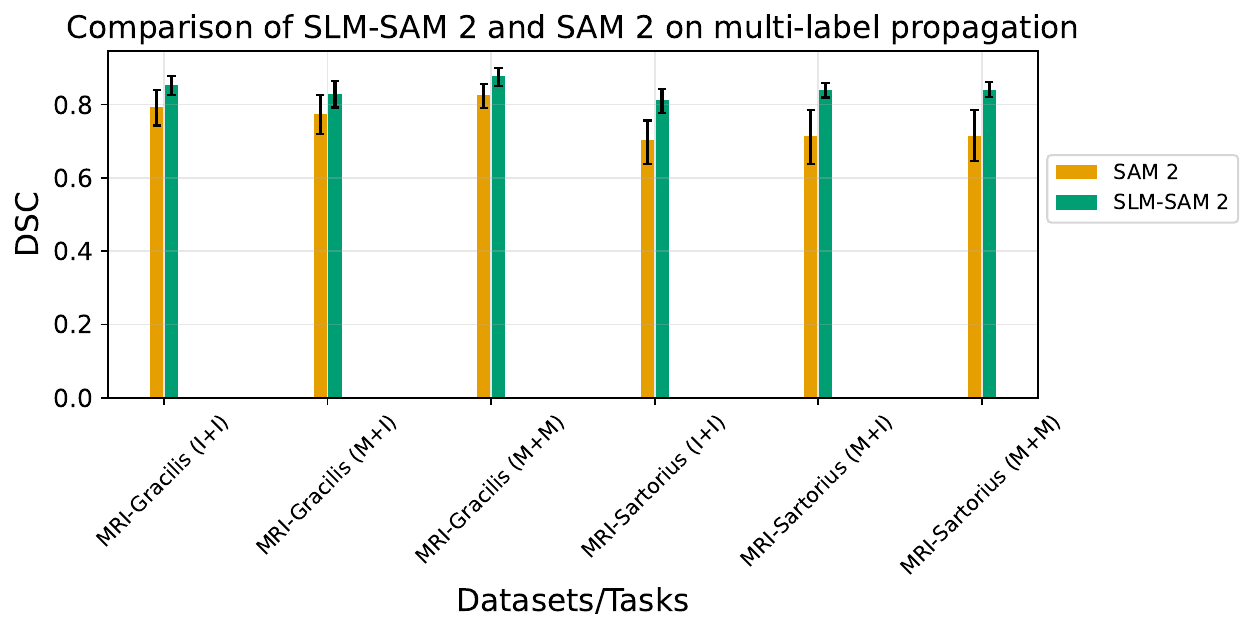}
    \caption{\textbf{Quantitative results of multi-label propagation.} SLM-SAM 2 consistently outperform SAM 2 in multi-label propagation.}
    \label{fig:multilabel_propagation}
\end{figure}

Fig. \ref{fig:multilabel_propagation} shows that SLM-SAM 2 outperforms SAM 2 across all three settings. Moreover, performance is comparable within each model across settings, demonstrating that a single finetuned SLM-SAM 2 can effectively propagate multiple classes.

\subsection{Annotation Time \& Correction Analysis}
\label{sec:annotation_time_corrections}
To reflect the practical value of SLM-SAM 2, we evaluated the efforts needed to correct annotations after label propagation. The assessment was performed on MRI-Kidney, CT-Femur and MRI-Sartorius, comparing SLM-SAM 2 to SAM 2 in terms of the \textbf{SPV} (seconds per volume) and the \textbf{CSR} (corrected-slice ratio), which is the number of corrected slices divided by the total number of slices. Two independent annotators, both non-physician researchers with substantial experience in medical image annotation, were blinded to which model produced each mask and corrected the labels on their assigned volumes in 3D Slicer \cite{fedorov20123d}. We then compute the DSC score between the corrected masks and ground truth (DSC$_{corr}$) to ensure the corrections reach equivalent quality. The average saved effort in SPV and CSR ratio are calculated as:

\begin{equation}
    SPV_{saved} = \frac{SPV_{SAM 2} - SPV_{SLM-SAM 2}}{SPV_{SAM 2}}
\end{equation}

\begin{equation}
    CSR_{saved} = \frac{CSR_{SAM 2} - CSR_{SLM-SAM 2}}{CSR_{SAM 2}}
\end{equation}

\begin{table}[!ht]
\centering
\caption{Comparison of practical annotation time and corrections needed between SAM 2 and SLM-SAM 2. SPV is seconds per volume and CSR is corrected-slice ratio.}

\resizebox{0.5\textwidth}{!}{%
\begin{tabular}{c|c|c|c|c}
\toprule
 \textbf{Datasets} & \textbf{Methods} & \textbf{SPV} & \textbf{CSR} & \textbf{DSC$_{corr}$} \\
\midrule
 MRI-Kidney & SAM 2 & 342.05 & 23.142 $\%$ & 0.9735 \\
 MRI-Kidney & SLM-SAM 2 & 203.40 & 17.128 $\%$ & 0.9744 \\
\midrule
 CT-Femur & SAM 2 & 515.10 & 42.191 $\%$ & 0.9585 \\
 CT-Femur & SLM-SAM 2 & 21.93 & 1.587 $\%$ & 0.9691 \\
\midrule
 MRI-Sartorius & SAM 2 & 134.05 & 44.371 $\%$ & 0.9470 \\
 MRI-Sartorius & SLM-SAM 2 & 96.70 & 36.755 $\%$ & 0.9588 \\
\midrule
\midrule
\multicolumn{2}{c|}{Avg. Saved Effort} & 60.575 $\%$ & 53.574 $\%$ & -- \\
\bottomrule
\end{tabular}
}
\label{tab:practical_time_corrections}
\end{table}

According to Table \ref{tab:practical_time_corrections}, SLM-SAM 2 reduces correction effort by an average of $60.575\%$ in SPV and $53.574\%$ in CSR compared to SAM 2. This is reasonable as SLM-SAM 2 produces more accurate initial masks than SAM 2, and thus, requires fewer manual corrections, underscoring the practical value of SLM-SAM 2. Notably, SLM-SAM 2 achieves a much lower SPV than SAM 2 on the CT-Femur dataset. This is because SAM 2 is more likely to over-propagate masks from the femur to the hip bone (as shown in Fig. \ref{fig:propagation}), which results in substantially more slices requiring correction for each volume, as reflected by its much higher CSR compared to SLM-SAM 2.

\subsection{Computational Efficiency Evaluation}
\label{sec:computational_efficiency}
We assessed the computational efficiency of SAM 2 and SLM-SAM 2 at inference by measuring \textbf{FPS} (frames per second) and peak GPU memory usage. Table \ref{tab:computational_usage} shows that SLM-SAM 2 exhibits slightly lower FPS and higher peak GPU usage, which is expected given its additional memory attention module and dynamic fuser.
\begin{table}[!ht]
\tiny
\centering
\caption{Comparison of computational usage between SAM 2 and SLM-SAM 2 at inference. FPS is frames per second.}

\resizebox{0.4\textwidth}{!}{%
\begin{tabular}{c|c|c}
\toprule
\textbf{Methods} & \textbf{FPS} & \textbf{Peak GPU Usage (GB)} \\
\midrule
 SAM 2 & 32.634 & 3.148 \\
 SLM-SAM 2 & 27.106 & 3.216 \\
\bottomrule
\end{tabular}
}
\label{tab:computational_usage}
\end{table}

\subsection{SLM-SAM 2 in 3D Slicer}
\label{sec:3d_slicer_extension}
3D Slicer is an open-source platform for medical image annotation \cite{fedorov20123d}. To integrate SLM-SAM 2 directly into this environment, we have developed a 3D Slicer extension (SegmentHumanBody) that enables interactive inference and mask propagation. Users can load finetuned SLM-SAM 2 checkpoint weights and propagate annotations in bi-directional mode. Generated segmentations may then be refined using 3D Slicer's built-in modules. The extension is available on GitHub at: \url{https://github.com/mazurowski-lab/SlicerSegmentHumanBody}

\section{Conclusion}
\label{sec:conclusion}

In this paper, we proposed SLM-SAM 2, a novel VOS method to accelerate volumetric medical image annotation by propagating annotations from a single slice to the remaining slices in volumetric medical data. SLM-SAM 2 introduces a dynamic short-long memory module that integrates both short-term and long-term memory, along with a learnable attention fuser. We evaluated SLM-SAM 2 on organs, bones, muscles and fetal head across four public datasets. Both quantitative and qualitative results have demonstrated the superior performance of SLM-SAM 2 compared to leading automatic 2D and 3D models, as well as state-of-the-art unsupervised and supervised VOS methods, effectively alleviating over-propagation issues and other common propagation errors. Furthermore, we showed that SLM-SAM 2 reduces manual correction time by $60.575\%$, highlighting its practical value in medical image annotation.

\bibliographystyle{IEEETran}
\bibliography{IEEEabrv,refs}
\end{document}